\documentclass[aps,prb,twocolumn,showpacs,groupedaddress]{revtex4}

\usepackage{amssymb}
\usepackage{amsmath}
\usepackage{dcolumn}
\usepackage{bm}
\usepackage{graphicx}
\usepackage{verbatim}

\providecommand*{\void}[1]{}

\begin{document}

\title{Scaling of 1/f noise in tunable break-junctions}

\author{ZhengMing Wu}
\author{SongMei Wu}
\author{S. Oberholzer}
\author{M. Steinacher}
\author{M. Calame}
\author{C. Sch{\"o}nenberger}

\email{christian.schoenenberger@unibas.ch}

\affiliation{Departement f{\"u}r Physik, Universit{\"a}t Basel, Klingelbergstr.~82, CH-4056 Basel, Switzerland }

\begin{abstract}
We have studied the $1/f$ voltage noise of gold nano-contacts in
electro\-migrated and mechanically controlled break-junctions
having resistance values $R$ that can be tuned from $10$\,$\Omega$
(many channels) to $10$\,k$\Omega$ (single atom contact). The
noise is caused by resistance fluctuations as evidenced by the
$S_V\propto V^2$ dependence of the power spectral density $S_V$ on
the applied DC voltage $V$. As a function of $R$ the normalized
noise $S_V/V^2$ shows a pronounced cross-over from $\propto R^3$
for low-ohmic junctions to $\propto R^{1.5}$ for high-ohmic ones.
The measured powers of $3$ and $1.5$ are in agreement with
$1/f$-noise generated in the bulk and reflect the transition from
diffusive to ballistic transport.

\end{abstract}







\date{\today}

\maketitle

\newpage
\section{Introduction}
%
The study of fluctuations (noise) in physical properties of condensed matter
has been an active area of research for decades and has led
to profound insights into time-dependent physical phenomena~\cite{Bak87,vdZiel54,Kogan96,Blanter2000,Beenakker2003}.
In case of charge transport, the noise shows up as a fluctuating
time-dependent AC-voltage $\delta V$ over the device with resistance $R$.
The most generic noise contributions stem from equilibrium thermal fluctuations
of the electron-bath (Johnson-Nyquist noise)~\cite{Johnson27,Nyquist28},
non-equilibrium shot noise caused by the granularity of charge~\cite{Schottky18}, and resistance
fluctuations~\cite{Hooge69,Voss76,Bell80,Hooge81,Dutta_Horn_81,Keshner82,Giordano83a,Giordano89,Weissmann88,Vandamme94}.
Whereas thermal and shot noise are frequency independent,
resistance fluctuations display a strong dependence which often
closely follows a $1/f$ relation over a large frequency range.
Because of the $1/f$ dependence, this noise contribution dominates
over thermal and shot-noise at lower frequencies.
$1/f$-noise has intensively been studied for bulk and thin film
conductors~\cite{Hooge69,Voss76,Bell80,Hooge81,Dutta_Horn_81,Keshner82,Giordano83a,Giordano89,Weissmann88,Vandamme94},
in particular as a diagnostic tool for the technologically
relevant electro\-migration mechanism~\cite{Sorbello91,Vandamme94a,Dagge96,Dong2006}.
Noise at low and high frequencies
has also been explored in small constrictions~\cite{Yanson82,Yanson84,Ralls91,Holweg92},
nano-electronic devices~\cite{Martinis92,Birk95}, quantum point-contacts~\cite{Liefrink92},
sub-micron interconnects~\cite{Neri97,Bid05}, quantum coherent, quasi-ballistic and
ballistic nanowires~\cite{Birge89,Strunk98,Birge99,Neuttiens2000,Collins2000,Snow2004,Fuhrer2006,Onac2006},
as well as tunneling contacts~\cite{Moeller89,Jiang90}.

The power-spectral density of resistance fluctuations $S_R$ can phenomenologically be
described by Hooge's law~\cite{Hooge81,Hooge69}:
\begin{equation}
  S_{R}(f)/R^2= \frac{\alpha}{Nf} \mbox{\ ,}
  \label{hooge}
\end{equation}
expressing proportionality of $S_R$ with a $1/f$ frequency dependence. The proportionality factor
$\alpha/N$ is ascribed to a material parameter $\alpha$ containing the strength of elastic and inelastic scattering,
on the one hand, and to an extensive variable $N$, on the other hand.
The constant $N$ denotes the number of statistically independent fluctuators in the volume.
It is straightforward to derive this $N$ dependence by assuming a resistance network with $N$ resistors in series
(or in parallel), all fluctuating independently. The total square fluctuation is then inversely proportional
to $N$.
In bulk conductors, the total number of electrons has been used for the
variable $N$~\cite{Hooge81,Giordano83a,Hooge90}. Partial support for this view comes from
semiconductors in which the carrier density can be changed over many orders
of magnitude~\cite{Hooge90,Tacano93,Vandamme2000}.
Hooge's law therefore states that $1/f$-noise is a bulk phenomenon, originating
homogeneously over the whole volume. In structures of reduced dimensionality, such as thin films and nano\-wires,
where the surface may dominate over the bulk, the leading contribution to $1/f$-noise may
stem from surface roughness and its fluctuations~\cite{Sah66,Hooge69,Vandamme89,Wong90}.
The validity of bulk scaling of $S_R$ has therefore been questioned. However, there are no quantitative
studies on the scaling behavior of $1/f$-noise in nano\-contacts with tunable cross sections in which
this dependence could be explored.

In this paper we report on $1/f$-noise measured in tunable metallic nano-constrictions
obtained through electro\-migration (EM)~\cite{Park1999} and mechanically
controlled break-junctions (MCBJs)~\cite{Ruitenbeek96,Scheer97,Agrait2003}.
Our emphasis is on the role of the scaling parameter $N$ in nano-contacts in the regime of few
transport channels where the transition from diffusive to ballistic transport takes place.
This transition is observed in our experiments at room temperature and we demonstrate
that even in nano\-contacts with only a few transport channels, $1/f$-noise is a bulk property.

\section{Experimental setup and calibration}

Representative examples of EM junctions and MCBJs are shown in Fig.~1a-b.
They are both fabricated using electron-beam lithography and metal deposition
in a lift-off process. In both cases Au wires with narrow constrictions with typical
dimensions of \mbox{$200$\,nm} in length and \mbox{$100$\,nm} in width are defined first.
Each wire has four terminals enabling the accurate measurement of the electrical resistance.
Au wires are fabricated on oxidized \mbox{($400$\,nm)} Si
substrates for EM junctions, and on a flexible substrate for
MCBJs, onto which a several $\mu$m thick insulating polyimide
layer is cast~\cite{Grueter2005}.
To form an EM junction the four terminals are used in an automatic feedback
controlled EM process which continuously shrinks the wire constriction
down to an atomic-sized nano\-junction as seen in Fig.~1a~\cite{Wu2007}.
In MJBJs the wire constriction is first transferred into a suspended bridge by etching the
underlying polyimide layer in an oxygen plasma as seen in Fig.~1b.
By bending the substrate the constriction can be narrowed in a
controlled manner~\cite{Ruitenbeek96,Scheer97,Agrait2003}.

Before narrowing the constrictions, the as-fabricated devices have
a junction resistance $R_J$ of around \mbox{$1-10$\,$\Omega$} at room temperature
as determined in a four terminal setup. The two-terminal resistance $R=R_J+2 R_L$, which
includes the lead resistance $R_L$ on both sides, amounts to as much as \mbox{$250$\,$\Omega$}.
Because $R_L>>R_J$ in virgin devices, the feedback-controlled process is
mandatory to initiate a nondestructive narrowing by EM~\cite{Wu2007,Trouwborst06}.
In voltage-biased controlled EM, in which the voltage over the
junction is stabilized by a fast analog feedback~\cite{Wu2007},
a narrowing sets in at a voltage of \mbox{$\agt 0.2$\,V}. The junction resistance $R_J$ then
rapidly evolves from a few Ohms to \mbox{$\approx 100$\,$\Omega$}. In this regime
of active EM, $R_J$ can further be increased into the
k$\Omega$-regime by increasing the junction voltage. An example of this process is shown in Fig.~1c.
We emphasize that we do not measure the noise while EM proceeds. After
narrowing the constriction at a `large' voltage we switch the applied voltage back
to values \mbox{$\alt 0.2$\,V}. During noise measurements, the junctions remain stable.
In contrast to EM junctions, MCBJs have the advantage that the junction size can
be changed with an independent control parameter by mechanical bending.
This allows to change the junction diameter
while monitoring noise simultaneously.

\begin{figure}[!htb]
\begin{center}
\includegraphics[width=8cm]{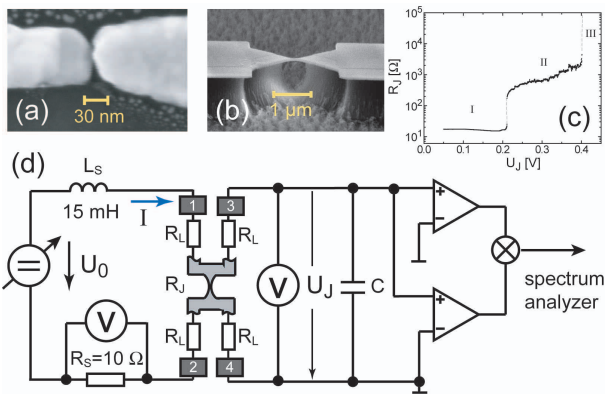} 
\end{center}
\caption{
  SEM images of (a) an EM junction and (b) a suspended
  bridge device used in the MCBJ setup. (c) shows a characteristic
  EM induced evolution of the junction resistance $R_J$ as a function of
  applied junction voltage $U_J$. Electro\-migration narrows the junction cross-section
  in regime {\rm II}~\cite{Wu2007}.
  (d) shows a schematics of the electric circuit used to measure the noise.
  $R_{L}$ denotes the resistance of each lead in the four-terminal setup and
  $C$ summarizes the total capacitance of the preamplifier and the connecting cables. The
  inductor $L_S$ is used to separate the AC-noise measurement on the right from the
  DC-biasing on the left.}
\end{figure}

We perform noise measurement in a four terminal setup schematically shown in Fig.~1d.
An adjustable low noise DC voltage source $U_0$ is connected via a series inductor $L_s$ and
a series resistor $R_s$ to contacts $1$ and $2$ on the left side, driving
a DC bias current $I$ through the junction. \mbox{$R_s=10$\,$\Omega$} is used to measure $I$
and \mbox{$L_s=15$\,mH} serves to decouple AC from DC. The impedance of the inductor prohibits
the shunting of the AC voltage fluctuations (noise). This only works if $\omega L_s > R_J$
($\omega$ is the angular frequency $=2\pi f$), defining a lower cut-off for the useful
frequency window.
The frequency-dependent noise is simultaneously
measured on terminal $3$ and $4$ and fed into two low-noise pre\-amplifiers (EG\&G5184)
and a spectrum analyzer (HP89410A). Here, the effective input capacitance $C$ is
diminishing the signal at high frequencies defining an upper cut-off for the frequency window
through the relation \mbox{$1/\omega C > R_J$}.
For a typical junction resistance $R_J$ of \mbox{$100$\,$\Omega$} and effective capacitance
\mbox{$C\approx 1$\,nF}, the useful frequency window spans approximately three
orders of magnitude, i.e. $1$\,kHz\,$< f < 1$\,MHz. We describe the $f$-dependence
of the circuit analytically (see below) and use this model to fit the total capacitance
which contains parts of the Si chip, the connecting wires and the amplifiers.
%
%
When measuring noise, the two preamplifiers measure the same fluctuating signal in parallel.
The spectrum analyzer is operated in the cross-spectrum mode and determines the Fourier transform
of the cross-correlation signal from the two amplifiers:
\begin{equation}
  S_V(f):=2\int_{-\infty}^{\infty} e^{i2\pi f t} \langle\delta U_1(t+\tau) \delta U_2(\tau) \rangle_{\tau} dt \mbox{\ ,}
  \label{Cross}
\end{equation}
where $\delta U_{1,2} (t)$ denotes the time dependent deviations from the average value of the junction voltage $U_J$
measured on amplifier $1$ and $2$, and $\langle \dots \rangle_{\tau}$ refers to averaging over $\tau$.
This signal is equivalent to the voltage power-noise spectral density. The correlation techniques
can eliminate the voltage noises originating from the two amplifiers because
the two amplifiers are independent.

All measurements are done at room temperature \mbox{($T=300$\,K)} and
thermal noise is used to calibrate the setup. The thermal noise of a resistor of value $R$
is given by $S_V=4kTR$. Due to the $f$-dependent elements in the circuit and the preamplifiers,
the noise signal is in general attenuated.
The attenuation factor $A$ has two components $A=A_1 \times A_2$.
$A_1$ is determined by the two circuits in Fig.~1d parallel to $R_J$, the one with the inductor $L$ on the left and
the one with the capacitor $C$ on the right. We obtain for this attenuation factor $A_1$:
\begin{equation}
  A_1 =\bigg\vert\frac{1}{1 + i\omega C R_J + 1/(i\omega L/R_J + R_{\Sigma}/R_J)} \bigg\vert^2 \mbox{\,,}
\end{equation}
where $R_{\Sigma}=2R_L+R_s$ and where we have assumed that $R_L << 1/\omega C$.
The second part $A_2$ is due to the frequency-dependent gain of the amplifiers.
We have carefully measured this dependence in between \mbox{$1$\,Hz}
and \mbox{$1$\,MHz} and found that the high-frequency roll-off can
accurately be modelled by a first-order low-pass filter with a cross-over
frequency of \mbox{$f_c=840$\,kHz}. Hence, $A_2$ is given by:
\begin{equation}
  A_2 =\bigg\vert\frac{1}{1 + i2\pi f/f_c}\bigg\vert^2 \mbox{\ .}
\end{equation}
All parameters $L$, $R_L$, $R_s$ and the overall gain can accurately be measured
except $C$. We therefore determine the capacitance $C$ by fitting the frequency
dependence of the measured thermal noise $S_V(f)$ to the expected value
$4kTR_J A_1(f) A_2(f)$. A consistent \emph{single} value of \mbox{$C=270$\,pF}
has been found for different junction resistances. The validity of this calibration procedure
is demonstrated in Fig.~2. In Fig.~2a the frequency dependence of the measured thermal noise
is shown for different metal-film calibration resistances $R$ ranging between \mbox{$10$\,$\Omega$}
to \mbox{$10$\,k$\Omega$} which were used instead of a real junction.
One can see that the $f$-dependence is very strong for large junction resistance values, whereas
a flat $f$-independent part is clearly visible in the opposite case.
The expected noise according to $4kT R A_1(f) A_2(f)$ is plotted as dashed curves
in Fig.~2a. A very good agreement with the measured noise is evident. In Fig.~2b we
display the corrected data, i.e. the measured noise divided by the attenuation factor $A$.
This procedure works very well in the shaded frequency window
over the whole resistance range as evidenced by the flat noise plateaus that coincide with
the expected thermal noise (horizontal lines). For the $1/f$ noise study we will therefore
restrict the frequency window to the shaded region of \mbox{$30<f<400$\,kHz}.

\begin{figure}[!htb]
\begin{center}
\includegraphics[width=8cm]{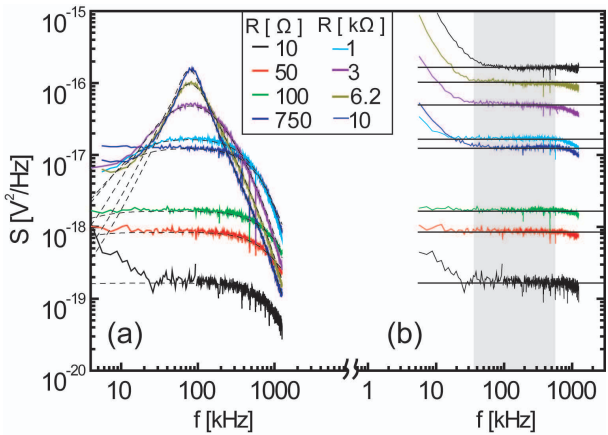} 
\end{center}
\caption{
  (a) Calibration of the setup by measuring the thermal noise of eight
  standard metal-film resistors with values $R$ ranging from \mbox{$10$\,$\Omega$} to
  \mbox{$10$\,k$\Omega$}. The dashed curves are calculated taking the
  frequency dependence of the circuit and gain of the amplifiers into account.
  In (b) the measured thermal noise is corrected for the known frequency
  dependence of the circuit and amplifier gain. The horizontal lines
  mark the theoretical thermal noise of $4kTR$.
  In the shaded region the corrected noise is frequency independent
  and coincide with the expected thermal noise values.
  This frequency interval is used to measure the $1/f$ noise
  in nano\-junctions.
}
\label{calibration}
\end{figure}
%

\section{Results and Discussion}
%
\begin{figure}[!htb]
\begin{center}
\includegraphics[width=8cm]{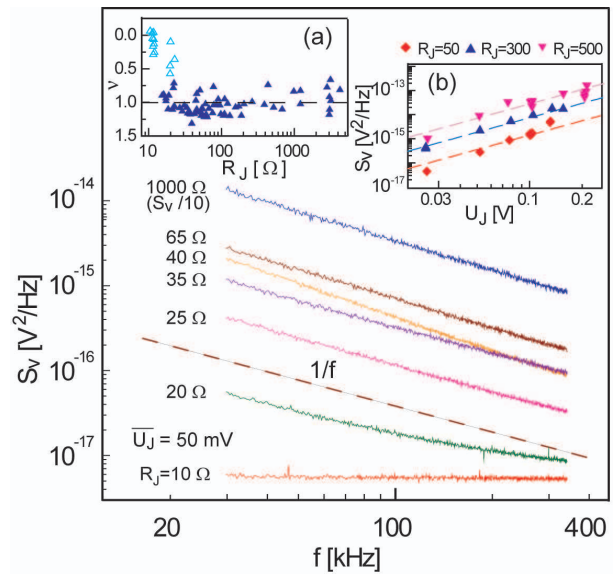} 
\end{center}
\caption{
  Noise spectra of a single EM junction, measured at
  \mbox{$U_{J}=50$\,mV} and for relatively low junction
  resistances $R_{J}$ ranging from $10$ to \mbox{$65$\,$\Omega$} (from bottom to top).
  For comparison we also show $S_V(f)$ for $R_J=1$\,k$\Omega$.
  Inset (a): The exponent $\nu$ of the frequency dependence deduced from the noise spectrum of many junctions
  with a large range of junction resistances $R_J$ and bias voltages $U_J$.
  The open symbols correspond to devices with small $R_J$ and measurements at
  small bias voltage $U_J$, displaying only white noise originating
  from the lead resistances. The average value of $\nu$ for the filled symbols
  is equal to $1.1$ and therefore close to the expected value for $1/f$-noise.
  Inset (b) shows the bias voltage dependence of $S_{V}$ at a frequency $f$
  of \mbox{$f_0=100$\,kHz} and for three different $R_J$ values.}
  \label{1/f}
\end{figure}

Figure~3 shows the $f$-dependence of $S_V(f)$ for a single EM
junction in the `low'-ohmic regime with $R_J=10\dots 65$\,$\Omega$ together
with a curve representative for large junction resistances (here, $R_J=1$\,k$\Omega$).
To measure $1/f$-noise we typically apply a voltage of $50$\,mV. This is below the threshold
for EM and enables the stable measurements of junctions.
Except for the lowest two curves, the main panel of Fig.~3 shows that $S_V(f)$ decays in a power-law
fashion, but this decay is not exactly inversely proportional to $f$.
This fact has often been noted before:
$S_V\propto 1/f^{\nu}$ with $\nu$ ranging between $1$ and $2$. The latter is expected
for a single two-level fluctuator~\cite{Holweg92}. In our case, the exponent $\nu$
is close to $1$ with an average of $\nu \approx 1.1$, taking all data
with $R_J> 20$\,$\Omega$. What is remarkable, however, is the
sample-to-sample fluctuation in the slope (particularly strongly visible in
the $R_J=40$\,$\Omega$ curve) which we observe universally in all devices.

In addition to the sample-to-sample fluctuation of the slope around a mean-value
of $\nu \approx 1.1$, we also see that the bottom curve for the smallest junction
resistance of \mbox{$R_J=10$\,$\Omega$} is flat and displays no $1/f$-noise.
This is also true for all devices: $1/f$-noise
only shows up for a sufficient large junction resistance $R_J$ and DC-bias
$U_J$. This is because the thermal noise of the series connection $R_J + 2 R_L$
dominates at a small bias $U_J$ and small $R_{J}$. After increasing $R_{J}$
at constant $U_J$, the $f$-dependence of $S_V$ sets in.
The $1/f$ dependence shows up first on the low frequency side. At the high frequency side
the thermal noise still dominates. This leads to the impression that the spectrum is flatter
than $1/f$ in this transition regime from `low' to `large' $R_J$ values.
The deduced power $\nu$ in the relation $S_V \propto 1/f^{\nu}$ is shown in the inset (a)
of Fig.~3. The open symbols belong to junctions with too low $R_J$ that do not
display $1/f$-noise in the given frequency interval and for applied voltage.
Only the filled symbols correspond to junctions displaying full $1/f$-noise.
There is quite some scatter in $\nu$,
but all values stay close to $\nu=1$.

In order to shed light on the origin of the $1/f$-noise, the voltage dependence of
$S_V$ has been analyzed. The second inset (b) of Fig.~3 shows $S_V$ taken at a fixed frequency of
\mbox{$100$\,kHz} as a function of $U_J$. The different symbols refer
to three representative samples with $R_J=50$, $300$ and \mbox{$500$\,$\Omega$}.
There is a strong increase of $S_V$ with $U_J$ which is in quite good
agreement with a quadratic dependence, i.e. $S_V\propto U_J^2$, for
not too large voltages \mbox{($\alt 0.2$\,V)}. This quadratic dependence agrees with our expectation
for resistance fluctuations as the source of $1/f$-noise. This expression can be understood by noting
that the fluctuating junction resistance $\delta R_J$ generates the fluctuating voltage
$\delta U_J=I\delta R_J$ over the junction at a constant DC bias current $I$.
The mean square fluctuation, i.e. the noise, is then proportional to $I^2$ and
therefore also to $U_J^2$.

Having established the $1/f$ dependence and confirmed resistance fluctuations as
its origin, we consider next the prefactor $\alpha/N$. $S_V$ of many samples has been
measured as a function of the junction cross-section, i.e. as a function
of $R_J$, and the $1/f$ contribution was extracted within the frequency interval
\mbox{$30-400$\,kHz} following the procedure described before. To compare the
magnitude of $S_V$ for different samples and different junctions, we now take
the normalized noise $S_V(f)/V^2$ at a fixed frequency of \mbox{$f=f_0=100$\,kHz}.

\begin{figure}[!htb]
\begin{center}
\includegraphics[width=8cm]{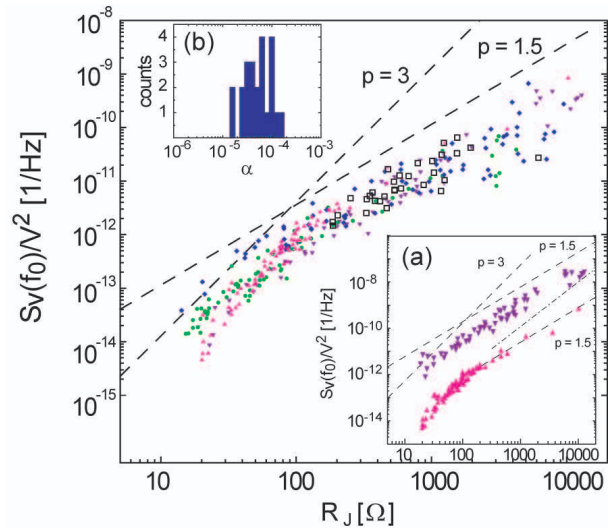} 
\end{center}
\caption{
    Magnitude of the $1/f$-noise, shown as a log-log scatter plot of $S_V(f_0)/V^2$ with $f_0=100$\,kHz
    of four MCBJ samples and one EM sample (open squares) as a function of junction resistance $R_J$.
    The inset (a) shows the data from two MCBJs separately where the upper set is vertically shifted by
    two-orders of magnitude for clarity.
    Dashed lines are guides to the eyes for the expected power-law dependencies $S_V(f_0)/V^2\propto R_J^p$
    in the diffusive ($p=3$) and ballistic ($p=1.5$) regime. The dashed-dotted line has a slope of $p=2$.
    Inset (b) shows a histogram of $\alpha$-values deduced from $S_V$ for the single-atom contact
    when $R_J=h/2e^2$.}
  \label{scaling}
\end{figure}

In Fig.~4 (main panel) we show a scatter plot of $S_V(f_0)/V^2$ as a function of $R_J$
of a few samples in a double-logarithmic representation.
Four sets were obtained with MCBJ samples and one with a EM one (open squares).
Note, that EM samples typically cover only the regime \mbox{$R_J \agt 100$\,$\Omega$},
because when EM sets in, there is a relatively fast transition from the low-ohmic
regime {\rm I} in Fig.~1c to the intermediate resistance regime {\rm II}.
The scatter plot clearly displays a cross-over from a power\-law dependence $S_V(f_0)/V^2\propto R_J^{\,p}$
with a large power for low $R_J$ and a smaller one for large $R_J$. This cross-over is better seen in inset (a).
Although there are some sample to sample variations, we always observe a cross-over in all of our samples
in the vicinity of \mbox{$R_J\approx 100$\,$\Omega$}. The deduced powers are consistent with $p=1.5$ and $p=3$
for large and low $R_J$, respectively. The transition and the deduced values are in agreement with
$1/f$-noise generated in the bulk together with a transition from the
diffusive to the ballistic transport regime with increasing $R_J$ as we will outline in the following.

It has been pointed out by Hooge~\cite{Hooge90}, that $1/f$-noise is a bulk phenomenon, whose scaling
parameter $N$ (see eq.~\ref{hooge}) should grow like the volume $\Theta$. Although this
has been disputed and was discussed many times over the last two decades, we will assume
scaling with volume and compare to scaling with the surface afterwards.
Let us denote a characteristic length of the junction by $l$. In order to refer to size-scaling, we use the
terminology length$\,\sim l$, which reads ``length scales with $l$''. Obviously, $\Theta \sim l^3$.
In a diffusive wire of length $L$ and cross-section $A$, the resistance $R$ is given by
$R=\rho L/A$, where $\rho$ is the specific resistance. Hence, $R\sim l^{-1}$. Because
$S_V/V^2\propto 1/N\sim l^{-3}$ (eq.~\ref{hooge}), we expect $S_V/V^2\propto R^3$ in this transport regime.
If the characteristic length of the junction becomes shorter than the momentum scattering mean-free path,
one is entering the ballistic regime. In this regime, the conductance is determined
by the number of transport channels which is proportional to the junction area. The corresponding junction resistance
is the so called Sharvin resistance~\cite{Sharvin65}. Hence, $R\propto A^{-1}\sim l^{-2}$.
Consequently, $S_V/V^2\propto R^{1.5}$.
The data in Fig.~4 shows a cross\-over which agrees with these derived exponents.

As a comparison, we also derive the expected power, if transport is ballistic and the fluctuators
leading to $1/f$-noise are only present on the surface. All transport channels in the interior
of the junction are assumed to be noise\-less. Now, $S_V/V^2$ will be inversely proportional
to $N_{S}$, where $N_S$ stands for the number of transport channels on the surface. This number
scales as the circumference, and therefore $N_S \sim l$. Using the Sharvin resistance for
a ballistic contact $R\propto A^{-1}\sim l^{-2}$ we arrive at $S_V/V^2\propto R^2$. Although
$p=2.0$ is not much different than $p=1.5$, we are able to distinguish between the two values.
In inset (a) of Fig.~4 the dash-dotted line corresponds to $p=2$. It is clear that the slope of
the measured data points is smaller proving that even in small metallic junctions, in which only a few
channels carry the charge current, all of them contribute to $1/f$-noise and not only the
channels close to the surface.

Finally, we can estimate the parameter $\alpha$ in eq.~\ref{hooge}. This parameter corresponds to the
noise value for $N=1$ at $f=1$\,Hz. If we associate with $N$ the number of electrons (which for
Au is the same as the number of atoms), we have to look at the $S_V$ value for the single atom
contact. Because $R_J$ is then given by the quantum resistance $h/2e^2\approx 13$\,k$\Omega$,
we find from Fig.~4 $S_V(f_0)/V^2 \approx 10^{-10}-10^{-9}$\,Hz$^{-1}$. Multiplying with $f_0=100$\,kHz yields
$\alpha \approx 10^{-5}-10^{-4}$. $\alpha$ values deduced in this way are shown as a histogram
in inset (b) of Fig.~4. This range of $\alpha$-values compares very well with parameters reported in the
literature~\cite{Holweg92}.

\section{Conclusions}

In conclusion, we have studied $1/f$-noise at room temperature in tunable electro\-migration and
mechanical controllable break junctions made from Au
in a regime in which only a few number of transport channels ($1-1000$)
contribute to the overall conductance. The transition from the diffusive to the ballistic transport
regime is clearly visible in the normalized noise $S_V/V^2$ when plotted against the junction
resistance $R_J$. This transition appears at \mbox{$R_J\approx 100$\,$\Omega$}. We find that
even in the smallest junctions, $1/f$-noise is a bulk property.

\section{Acknowledgement}
%
This work has been supported by the Swiss National Center (NCCR) on
"Nanoscale Science", the Swiss National Science Foundation, and the
University of Basel.

\end{document}